\documentclass{llncs}

\usepackage[utf8]{inputenc}
\usepackage{fontenc}
\usepackage{url}
\usepackage{textcomp}
\usepackage{listings}
\usepackage{graphicx}

\begin{document}

\title{L-FLAT: Logtalk Toolkit for\\Formal Languages and Automata Theory}
\titlerunning{L-FLAT: Logtalk Toolkit for Formal Languages and Automata Theory}

\author{Paulo Moura\inst{1} \and Artur Miguel Dias\inst{2}}
\authorrunning{Paulo Moura \and Artur Miguel Dias}

\institute{Dep. of Computer Science, University of Beira Interior, Portugal\\
           Center for Research in Advanced Computing Systems, INESC Porto, Portugal\\
           \email{pmoura@di.ubi.pt}
           \and 
           CITI / Dep. of Computer Science, Faculdade de Ciências e Tecnologia\\
           Universidade Nova de Lisboa, Portugal\\
           \email{amd@di.fct.unl.pt}
}

\maketitle

\begin{abstract}
We describe L-FLAT, a Logtalk Toolkit for teaching Formal Languages and Automata Theory. L-FLAT supports the definition of \textsl{alphabets}, the definition of \textsl{orders} over alphabet symbols, the partial definition of \textsl{languages} using unit tests, and the definition of \textsl{mechanisms}, which implement language generators or language recognizers. Supported mechanisms include \textsl{predicates}, \textsl{regular expressions}, \textsl{finite automata}, \textsl{context-free grammars}, \textsl{Turing machines}, and \textsl{push-down automata}. L-FLAT entities are implemented using the object-oriented features of Logtalk, providing a highly portable and easily extendable framework. The use of L-FLAT in educational environments is enhanced by supporting Mooshak, a web application that features automatic grading of submitted programs.\\

\end{abstract}

\section{Introduction}

L-FLAT \cite{lflat} is a free, open source, Logtalk \cite{pmoura03,lgtuserman2400} Toolkit for teaching Formal Languages and Automata Theory. It supports the definition of \textsl{alphabets} and total \textsl{orders} over alphabet symbols, the partial definition of \textsl{languages} using unit tests, and the definition of \textsl{mechanisms}, which attempt to implement a particular language, using either a language generator or a language recognizer. Supported mechanisms include \textsl{predicates}, \textsl{regular expressions}, \textsl{finite automata}, \textsl{context-free grammars}, \textsl{push-down automata}, and \textsl{Turing machines}.

As an educational tool, L-FLAT implements algorithms and concepts taught in typical courses of Formal Languages and Automata Theory (FLAT). These include algorithms for recognizing and generating words, for converting between different types of mechanisms, for querying different aspects of the mechanisms, for checking if a mechanism is deterministic, for making finite automata deterministic, and for minimizing mechanism definitions. L-FLAT is designed primarily as an interactive tool with a command-line interface but its use in educational environments is enhanced by supporting Mooshak, a web application that features automatic grading of submitted programs. To maximize its usefulness, L-FLAT is implemented as a highly portable, easily extensible tool.

The remainder of the paper is organized as follows. Section 2 briefly characterizes Logtalk as a programming language. Section 3 presents an extended tutorial example, illustrating some of the main L-FLAT features. Section 4 gives an overview of the current L-FLAT implementation. Section 5 describes typical classroom usage scenarios for L-FLAT. Section 6 discusses the L-FLAT --- Mooshak integration. Section 7 compares L-FLAT with related work. Section 8 describes the current implementation status and outlines future work.

\section{Lotalk in a Nutshell}

L-FLAT is implemented in Logtalk, an object-oriented logic programming language that can use most Prolog implementations as a back-end compiler. Logtalk focus in code encapsulation and code reuse features, providing an alternative to Prolog module systems. Logtalk supports classes, prototypes, parametric objects, categories (fine-grained units of code reuse), and separation between interface and implementation using protocols. We take advantage of all these features in L-FLAT. Logtalk uses \textsl{object} as a generic term: an object can play the role of e.g. an instance, a class, or a prototype. The relations between objects, protocols, and categories defining different \textsl{patterns of code reuse}. Logtalk entities can be static, defined in source files, or dynamic, created at runtime. Computations are performed by sending \textsl{messages} (corresponding to predicates) to objects. Logtalk enforces predicate encapsulation (predicates can be declared public, protected, or private) and features a clear distinction between predicate declaration and predicate definition (using a closed-world assumption when a predicate is declared but not defined). 


\section{A Tutorial Example}

In this tutorial, we will use the language comprised of all sequences of \textsl{a}s and \textsl{b}s with an even number of \textsl{b}s as a running example. We start with a partial language definition, followed by six different mechanisms, acting as recognizers or generators for the language. We end this section with examples of defining alphabets and alphabet orders. All source code is included in the current L-FLAT distribution.

\subsection{Language Definitions}

Languages are \textsl{partially} defined by stating three elements: their alphabet, a set of positive unit tests, and a set of negative unit tests. Listing \ref{evenL} provides an object definition for our example language.  Each unit test defines a word, a sequence of alphabet symbols (represented as a list), that should be either recognized or rejected by a language mechanism.

\begin{lstlisting}[caption={Sequences of \textsl{a}s and \textsl{b}s with an even number of \textsl{b}s}, label=evenL]
:- object(evenL,
    instantiates(language)).

    alphabet([a,b]).

    positive([]).
    positive([a,a,a]).
    positive([b,b]).
    positive([a,b,b]).
    positive([b,b,a,a]).

    negative([b]).
    negative([b,b,b]).
    negative([b,a,a]).
    negative([a,a,b]).
    negative([b,a,a,b,b,a,a]).

:- end_object.
\end{lstlisting}

\noindent
Teachers often help the students define an initial set of unit tests while keeping a more comprehensive set of unit tests private for grading. A \lstinline{test_mechanism/1} predicate, defined in the \lstinline{language} class, can be used by students to test their mechanism definitions against a language unit tests (see Listing \ref{evenP} for an usage example of this predicate).

\subsection{Mechanism definitions}

A mechanism attempts to implement a specific language, acting as a \textsl{language generator}, as a \textsl{language recognizer}, or both. As the names suggest, a language generator can generate words belonging to a specific language while a language recognizer can recognize if a word belongs to a specific language.

\subsubsection{Predicate Mechanisms}

The most basic mechanism that we can define is a \textsl{predicate}. Listing \ref{evenP} provides an example. Predicate mechanisms should provide a suitable definition for the object predicate \lstinline{accept/1}, used for recognizing language words. The \lstinline{evenP} object defines this predicate by using the services of the object \lstinline{word}, which provides a set of utility predicates for constructing, decomposing, and sorting words.

\begin{lstlisting}[caption={Predicate mechanism for the \lstinline{evenL} language}, label=evenP]
:- object(evenP,
    instantiates(predicate)).

    :- initialization(
        evenL::test_mechanism(evenP)
    ).

    alphabet([a,b]).

    accept(Word) :-
        alphabet(Alphabet),
        word::word_alphabet(Word, Alphabet) ->
        word::occurs(b, Word, N), 0 =:= N mod 2.

:- end_object.
\end{lstlisting}

\noindent
The \lstinline{evenP} object also defines an initialization goal for testing the mechanism at loading time using the unit tests of the \lstinline{evenL} language defined in Listing \ref{evenL}.

\subsubsection{Regular Expression Mechanisms}

Listing \ref{evenRE} shows a regular expression for the \lstinline{evenL} language. Regular expressions are defined using the object predicate \lstinline{expression/1}. L-FLAT uses the infix operators \lstinline{+} and \lstinline{*} for expressing alternation and concatenation, and uses the postfix operators \lstinline{^*} and \lstinline{^+} for expressing repetition of a sub-expression.

\begin{lstlisting}[caption={Regular expression for the \lstinline{evenL} language}, label=evenRE]
:- object(evenRE,
    instantiates(re)).

    :- initialization((
        evenL::test_mechanism(evenRE),
        ::fa(FA),
        FA::determine(FAD),
        FAD::minimise(FAM),
        FAM::rename(FAR),
        FAR::show
    )).

    expression((a + b * a^* * b)^*).

:- end_object.
\end{lstlisting}

\noindent
The initialization goal of the \lstinline{evenRE} object illustrates how to convert a regular expression into a deterministic finite automaton. For educational purposes, each step of the conversion is implemented by a different predicate. Upon loading of the \lstinline{evenRE} object, the output in Listing \ref{convert} is generated.


\begin{lstlisting}[caption={Output generated when loading the \lstinline{evenRE} regular expression}, label=convert]
Starting tests of evenRE against evenL ...
... tests finished

FINITE AUTOMATON:
  fa(s1, [s1/a/s1,s1/b/s2,s2/a/s2,s2/b/s1], [s1])
  Initial state: s1
  Transitions:
	s1 - a -> s1
	s1 - b -> s2
	s2 - a -> s2
	s2 - b -> s1
  Final states: [s1]
  Deterministic: yes
\end{lstlisting}

\noindent
The generated automaton, represented using a \lstinline{fa/3} compound term, is a Logtalk \textsl{parametric object proxy} \cite{pmoura11a}, i.e. an \textsl{instantiation} of the identifier of the parametric object \lstinline{fa(Initial, Transitions, Finals)}, defined by L-FLAT (see next section). Parametric object proxies provide a compact representation for simple objects and are often used to represent the result of mechanism conversion predicates.

L-FLAT mechanisms understand a \lstinline{word/1} message for generating language words, as illustrated in the example query in Listing \ref{word}.

\begin{lstlisting}[caption={Generating \lstinline{evenL} words using the \lstinline{evenRE} regular expression}, label=word]
| ?- evenRE::word(Word).

Word = [] ;
Word = [a] ;
Word = [a, a] ;
Word = [b, b] ;
Word = [a, a, a] ;
Word = [b, b, a] ;
...
\end{lstlisting}

\subsubsection{Finite Automaton Mechanisms}

Listing \ref{evenFA} shows a finite automaton definition for the \lstinline{evenL} language. The transitions between states are expressed as \lstinline{FromState/Symbol/ToState} compound terms.

\begin{lstlisting}[caption={Finite automaton for the \lstinline{evenL} language}, label=evenFA]
:- object(evenFA,
    instantiates(fa)).

    initial(1).
    transitions([1/a/1, 1/b/2, 2/a/2, 2/b/1]).
    finals([1]).

:- end_object.
\end{lstlisting}

\noindent
Any L-FLAT mechanism may alternatively be represented as a compound term, that can be interpreted as a parametric object proxy, as illustrated in the previous section. In fact, all L-FLAT mechanisms, represented by classes, have a corresponding parametric object instance. Listing \ref{parametric} shows the definition of the \lstinline{fa/3} parametric object.

\begin{lstlisting}[caption={Parametric object \lstinline{fa/3}}, label= parametric]
:- object(fa(_Initial, _Transitions, _Finals),
	instantiates(fa)).

	initial(Initial) :-
		parameter(1, Initial).
	transitions(Transitions) :-
		parameter(2, Transitions).
	finals(Finals) :-
		parameter(3, Finals).

:- end_object.
\end{lstlisting}

\noindent
Parametric objects and object proxies provide an alternative representation for simple mechanisms and greatly simplify both L-FLAT internal tasks (e.g. conversion between mechanisms, bypassing the need of dynamically creating new objects at runtime) and student interaction with L-FLAT (e.g. entering a mechanism definition at the interpreter prompt).

\subsubsection{Context-free Grammar Mechanisms}
Listing \ref{evenCFG} shows an example of a context-free grammar for the \lstinline{evenL} language. Grammar rules are represented as \lstinline{Head->Body} where \lstinline{Head} is a non-terminal symbol and \lstinline{Body} is a (possibly empty) list of symbols.
\begin{lstlisting}[caption={Context-free grammar for the \lstinline{evenL} language}, label=evenCFG]
:- object(evenCFG,
    instantiates(cfg)).

    start_symbol('S').
    rules([
        ('S' -> [a, 'S']),
        ('S' -> [b, 'S', b]),
        ('S' -> ['S', 'S']),
        ('S' -> [])
    ]).

:- end_object.
\end{lstlisting}

\subsubsection{Pushdown Automaton Mechanisms}
Listing \ref{evenPDA} shows an example of a pushdown automaton for the \lstinline{evenL} language. The state transitions are represented as \lstinline{SourceState/PoppedSymbol/InputSymbol/TargetState/PushedSymbols} compound terms.

\begin{lstlisting}[caption={Pushdown automaton for the \lstinline{evenL} language}, label=evenPDA]
:- object(evenPDA,
    instantiates(pda)).

    initial(p).
    initial_stack_symbol(z).
    transitions([
        p/z/a/p/[z],
        p/z/b/q/[z],
        q/z/a/q/[z],
        q/z/b/p/[z]
    ]).
    finals([p]).

:- end_object.
\end{lstlisting}

\subsubsection{Turing Machine Mechanisms}

In Listing \ref{evenTM} we show an example of a Turing machine for the \lstinline{evenL} language. Each state transition is represented as a \lstinline{SourceState/ReadSymbol/} \lstinline{WrittenSymbol/MoveSymbol/TargetState} compound term.

\begin{lstlisting}[caption={Turing machine for the \lstinline{evenL} language}, label=evenTM]
:- object(evenTM,
    instantiates(tm)).

    initial(q0).
    transitions([
        q0/'B'/'B'/'R'/q1,
        q1/a/a/'R'/q1,
        q1/b/b/'R'/q2,
        q1/'B'/'B'/'R'/q3,
        q2/a/a/'R'/q2,
        q2/b/b/'R'/q1,
        q2/'B'/'B'/'R'/q4
    ]).
    finals([q3]).

:- end_object.
\end{lstlisting}

\noindent
L-FLAT provides a diagnostics predicate, \lstinline{diagnostics/0}, that can be used by the student to check mechanisms definitions for possible problems. Listing \ref{diagnostics} shows the output of this predicate for the \lstinline{evenTM} Turing machine.

\begin{lstlisting}[caption={Running \lstinline{diagnostics/0} on the \lstinline{evenTM} Turing machine}, label=diagnostics]
| ?- evenTM::diagnostics.

Starting diagnostics of evenTM ...
Warning in evenTM:
  undefined transition for state/symbol q0/a
Warning in evenTM:
  undefined transition for state/symbol q0/b
Warning in evenTM:
  undefined transition for state/symbol q3/a
Warning in evenTM:
  undefined transition for state/symbol q3/b
Warning in evenTM:
  undefined transition for state/symbol q4/a
Warning in evenTM:
  undefined transition for state/symbol q4/b
... diagnostics finished
yes
\end{lstlisting}

\noindent
L-FLAT allows the user (usually the teacher) to define the level of detail of error messages. Warnings during the compilation of entities (e.g languages and mechanisms) can be turned off, set to provide minimal information (e.g. when using Mooshak for student grading), or set to provide detailed information. Error messages (e.g. when testing a mechanism using the corresponding language unit tests) can also be configured to either reveal or conceal the cause of the problem.

L-FLAT mechanisms also accept a \lstinline{tracing/1} message that allows students to trace the parsing of a language word. An example using the \lstinline{evenTM} Turing machine is presented in Listing \ref{tracing}.

\begin{lstlisting}[caption={Tracing  parsing a \lstinline{evenL} word using the \lstinline{evenTM} Turing machine}, label=tracing]
| ?- evenTM::tracing([a,b,b,a]).

TRACING TURING MACHINE:
  Name: evenTM
  Traced word: [a, b, b, a]
  At least one execution path stops at an
  acceptance state.
  Word accepted.
  Traced steps:
    >B a b b a B	q0
     B>a b b a B	q1
     B a>b b a B	q1
     B a b>b a B	q2
     B a b b>a B	q1
     B a b b a>B	q1
     B a b b a B>B	q3
yes
\end{lstlisting}

\subsection{Alphabets and Orders}

L-FLAT supports the definition of alphabets by specifying an expression that returns the list of alphabet symbols, as exemplified in Listing \ref{bits}.

\begin{lstlisting}[caption={A simple alphabet, \lstinline{bits}, for a binary language}, label=bits]
:- object(bits,
	instantiates(alphabet)).

	expression([0,1]).

:- end_object.
\end{lstlisting}

\noindent
It is also possible to compose alphabets. For example, Listing \ref{decimal} illustrates how to define a new alphabet, \lstinline{decimal}, by reusing the \lstinline{bits} alphabet.

\begin{lstlisting}[caption={Defining a new alphabet, \lstinline{decimal}, by reusing another alphabet}, label=decimal]
:- object(decimal,
	instantiates(alphabet)).

	:- initialization((
		::show,
		::diagnostics
	)).

	expression(bits + [2,3,4,5,6,7,8,9]).

:- end_object.
\end{lstlisting}

\noindent
Listing \ref{up} defines an order over the \lstinline{bits} alphabet symbols. Alphabet orders can be used to perform lexicographic sorting of language words. We can define as many orders as needed over the same alphabet. The default order is expressed by the ordering in the list of symbols returned by alphabet expressions.

\begin{lstlisting}[caption={Defining an order over the symbols of alphabet \lstinline{bits}}, label=up]
:- object(up,
	instantiates(order)).

	alphabet(bits).
	sequence([0,1]).

:- end_object.
\end{lstlisting}

\noindent

\subsection{Playing With Words}

L-FLAT includes an utility object, \lstinline{word}, providing a set of predicates for e.g. computing words from regular expressions, generating words from an alphabet, generating words giving a specific word an order, and decomposing words into prefixes, suffixes, and sub-words. Listing \ref{wordex} shows an usage example, using the alphabets and orders defined above (in the first query, \lstinline{w^n} means \lstinline{w} repeated \lstinline{n} times, when \lstinline{n >= 0}; \lstinline{w^(-1)} means the reverse of \lstinline{w}).

\begin{lstlisting}[caption={Using the \lstinline{word} utility predicates}, label=wordex]
| ?- word::compute_word([0,1]^2*[1,1,0]^(-1), Word).
[0,1,0,1,0,1,1]
yes

| ?- word::word_alphabet(Word, bits).
[] ;
[0] ;
[1] ;
[0,0] ;
...

| ?- word::lexically_ordered(Word, [1,1,1], up).
[]
[1]
[1,1]
[1,1,0|_G1399]
[1,0|_G1396]
[0|_G1393]
yes

| ?- word::mixed_ordered(Word, [1,1,1], up).
[]
[_G1407]
[_G1407,_G1410]
[1,1,0]
[1,0,_G1413]
[0,_G1410,_G1413]
yes

| ?- word::next_word([1,1,1], Word, up).
[0,0,0,0]
yes

| ?- word::subword(Word, [1,2,3,4,5]).
[] ;
[1] ;
[] ;
[1,2] ;
...
\end{lstlisting}

\section{Implementation}

Alphabets, orders, languages, and mechanisms are implemented by taking advantage of the object-oriented features of Logtalk. All these kinds of entities are implemented as sub-classes of a generic L-FLAT \lstinline{entity} root abstract class, thus sharing a common protocol for printing, validating, and diagnosing possible problems. Mechanisms are implemented using a class hierarchy, allowing easy sharing of common code using inheritance. Concrete alphabets, orders, languages, and mechanisms can be defined either as instances of these classes or as Prolog facts interpreted as parametric object proxies by Logtalk, depending on their complexity. The current implementation has 4000 lines of Logtalk source code (including basic object and predicate documentation\footnote{This documentation can be automatically converted in HTML or PDF, providing a reference manual for using and extending L-FLAT.}), defining 19 classes, some prototypes (\textsl{stand-alone} objects, such as \lstinline{word}, encapsulating utility predicates), some protocols (interfaces), 16 parametric objects (necessary to support object proxies), and 3 objects dedicated to user interaction and Mooshak support. Figure \ref{arch} in the Appendix depicts the main L-FLAT entities and the main predicates implemented by these entities\footnote{A full entity diagram is included in the current L-FLAT distribution and can also be found here: \url{http://code.google.com/p/lflat/source/browse/trunk/lflat_entity_diagram.pdf}}


The recognition and generation operations - (\lstinline{accept/1}, \lstinline{accept/2} and \lstinline{word/1}) - are semi-algorithms, i.e. procedures not ensuring termination. They are implemented in a generic way inside the \lstinline{mechanism} abstract class, which is a super-class of all the L-FLAT mechanisms. These generic definitions are inherited and used without modification by the finite automata, the push-down automata, and the Turing machines. However, regular expressions and context-free grammars provide specialized definitions that ensure termination. In the case of context-free grammars, we use the relatively advanced but well known technique of converting each context-free grammar into its quasi-lambda-free and chain-free form.

A challenging task was coping with the non-termination of the semi-algorithms. Our solution was to use the breadth-first search strategy with loop detection: we build in parallel all the execution paths that may lead to success (called viable execution paths) and prune any loop once detected. This completely solves the problem for finite automata. However, non-termination remains an issue for those push-down automata and Turing machines with the property that none of the execution paths will ever reach success but there is at least one execution path endlessly generating distinct machine configurations, never repeating a machine configuration. In this situation the program runs out of memory or reaches a timeout, causing an exception to be raised and caught; the word being tested is neither accepted nor rejected.

Most of the L-FLAT code is based on the formal concepts and algorithms taught in classes, translated as straightforwardly as possible into Logtalk. Therefore, in general, the students should be able to examine the L-FLAT source code and recognize the concepts and algorithms they have learned in classes.

L-FLAT is available under the GNU General Public License v2 and can be downloaded from the Google Code web site. The current distribution includes a set of examples based on teaching textbooks \cite{monteiro03}.

\section{L-FLAT in the Classroom}

Although L-FLAT is a recent development, our past experience with its precursor, P-FLAT \cite{wermelinger+dias,pflat}, provides us with valuable insight on typical and successful classroom usage scenarios. L-FLAT is designed as an interactive tool that helps students perform experiments when learning formal languages and automata theory. The partial definition of a language from an alphabet and a set of positive and negative unit tests allows the student to interactively test his/her language recognizers (generators), tracking language words that are not properly recognized (generated) or wrongly recognized (generated). The mechanism testing informative warnings and error messages and the mechanism tracing features helps the student to diagnose and correct wrong definitions. L-FLAT can also play an important role in student self-assessment. Given a formal language specification or an informal description, the student can define unit tests that represent his/her understanding of which words belong and which words don't belong to the language, and implement one or more supported mechanisms, testing them against the partial language definition. The use of L-FLAT in the classroom, both in student self-assessment and teacher grading tasks, is further enhanced by its integration with Mooshak, described next.

\section{Mooshak Integration}

Mooshak \cite{SPE03} is a web-based tool for managing programming contests, with automatic judging, often used in undergraduate programming contests. It is also increasingly being used as a teaching tool with several goals: to drive programming related exercises, to pre-evaluate assignments, and more simply to receive and validate assignment submissions without any pre-evaluation. In practice, the students tend to interact with Mooshak from home, although using it in the classroom is also possible.

Mooshak works by accepting, compiling, and running program submissions, comparing the resulting output against the expected output, and giving immediate feedback. Multiple case tests are used to evaluate the correction of each solution. The time and space complexity of the solution can also be assessed by defining computation time and memory usage limits when setting up a contest.

L-FLAT is distributed with a compatibility layer for allowing its use by Mooshak. A demo Mooshak contest, named \lstinline{L-FLAT Demo} and consisting of ten problems, is available from the L-FLAT web site \cite{lflat}.

The Mooshak interface is attractive for students due to its simplicity. There is no need to install and setup Logtalk and L-FLAT or to study L-FLAT and its commands. The student simply reads the statement of each problem, writes a solution following the syntax of a model example, submits the solution, and gets instant feedback. This user experience is necessarily limited when compared with using L-FLAT directly. However, when the use of L-FLAT is not mandatory in a course, the Mooshak interface is probably the best way to drive most students to take advantage of L-FLAT, even if in a simplified way.

Another point is that using L-FLAT through Mooshak is somewhat compelling, and has some potential to increase the average student's interest for the course, assuming that it is available for a long running contest that includes most of the exercises of the course. The general experience with such a contest feels like an interactive game of solving puzzles, and there is a motivation for trying to reach higher and higher scores.


We end this section with an example illustrating how L-FLAT works with Mooshak. Consider the following problem statement:

\begin{quote}
Please create a context-free grammar, ``evenCFG'', that generates the language of the sequences of ``a''s and ``b''s, where the symbol ``b'' appears an even number of times. Do not forget that zero is an even number.
\end{quote}

\noindent
There are infinitely many solutions for this problem. Suppose that the student submits the solution found in Listing \ref{evenCFG}, which is correct. Responding to this submission, Mooshak runs a command that starts Logtalk and loads the L-FLAT application and the submitted object.

In our demo contest there is one single input test file per problem as an L-FLAT \lstinline{language} instance can include multiple unit tests. The content of the input test file is the \lstinline{evenL} object found on Listing \ref{evenL}, with the following initialization directive added:

\begin{lstlisting}[caption={\lstinline{evenL} language initialization for testing the \lstinline{evenCFG} mechanism}, label=initialization]
:- initialization((
    contests::diagnostics(evenL),
    contests::check_definition(cfg, evenCFG),
    contests::diagnostics(evenCFG),
    contests::test_mechanism(evenL , evenCFG),
    contests::finish_checking
)).
\end{lstlisting}

\noindent
This initialization directive allows the test to auto-run upon loading of the \lstinline{evenL} object. The initialization goal performs a basic validation of the \lstinline{evenCFG} context-free grammar and checks it against the positive and negative examples defined on the \lstinline{evenL} language. As the particular submission we are considering is correct, the generated output
matches exactly the contents of the output test file:

\begin{lstlisting}[caption={Output test file contents for checking the \lstinline{evenCFG} mechanism}, label=output]
Starting diagnostics of evenL ...
... diagnostics finished
evenCFG is well defined
Starting diagnostics of evenCFG ...
... diagnostics finished
Starting tests of evenCFG against evenL ...
... tests finished
Finished checking
\end{lstlisting}

\noindent
Thus, Mooshak accepts the submission. Any ill-formed submission or any well-formed submission incompatible with the unit tests would be rejected by Mooshak as a consequence of the extra messages generated.

\section{Related Work}

L-FLAT development was bootstrapped by a full Logtalk rewrite of P-FLAT, a Prolog Toolkit for teaching Formal Languages and Automata Theory. This rewrite, motivated by increasing maintenance problems of P-FLAT due to its growing code base, has important advantages over the original application while retaining and expanding its features. First, much improved compatibility with Prolog compilers, inherited from the Logtalk portability itself.\footnote{Using Logtalk 2.40.1, L-FLAT runs as-is on B-Prolog, Ciao Prolog, CxProlog, ECLiPSe, GNU Prolog, Qu-Prolog, SICStus Prolog, SWI-Prolog, XSB, and YAP.} The original P-FLAT implementation used Prolog features only available in some compilers, severely restricting its portability. Second, the refactoring of the original source code lead to a significant reduction of the number of arguments of most predicates, resulted in smaller and simpler predicate definitions. Despite the fact that P-FLAT worked with general concepts (e.g. Mechanism) and specializations of those concepts (e.g. Turing machine), it was implemented as a plain Prolog application. The low level solution used for simulating inheritance and code sharing within the application required the presence of extra arguments in most predicates and a complex machinery just to keep track of entity identifiers (around 8\% of the original code). Third, the consequent code modularization further makes the code easier to understand, improves reliability, and simplifies future extension of the application, by its current developers or by external contributors. For example, we recently added support for context-free grammars and Turing machines, which were easily implemented as sub-classes of the existing \lstinline{mechanism} abstract class, thus inheriting useful features already in-place common to all mechanisms. In the old P-FLAT application, adding a mechanism required widespread and thus more error-prone changes to the code base.

When compared with related educational tools, L-FLAT innovates by allowing languages to be partially defined using unit tests, which can then be used to test the formal language mechanisms. This feature increases the ability of the students to use L-FLAT as an exploratory tool. Moreover, the combination of language unit tests with the textual and declarative definition of mechanisms provides the necessary support for using L-FLAT with Mooshak, further enhancing its pedagogical value.

The FSA Utilities toolbox \cite{fsa} is an open-source Prolog implementation of a set of utilities for working with regular expressions, finite-state automata, and finite-state transducers. This toolbox implements several algorithms for constructing finite automata from regular expressions and for manipulating finite automata. Although not as portable as L-FLAT, the FSA toolbox includes a SICStus Prolog specific graphical user interface, besides a command-line interpreter. It is also possible to compile finite automata as stand-alone executables (using the C programming language). The FSA toolbox can also be used as an education tool (extensive course material is available from the software web page \cite{fsaweb}), as a command-line filter utility, or as a general purpose Prolog library.

Similar educational tools exist written in other programming languages. The best know example is  JFLAP \cite{jflap}, an application written in Java. JFLAP provides a set of graphical tools for visually designing and simulating different mechanisms. However, it does not support user-written textual definitions of mechanisms as found in L-FLAT. Although the current version of L-FLAT only provides a textual interface, it provides some significant advantages: declarative definition of alphabets, orders, languages, and mechanisms; user-defined operators, which improve readability; built-in unification and non-determinism, making it simple to define and explore non-deterministic mechanisms; inherent interactive nature of the Logtalk/Prolog interpreter, which favors experimentation by students learning FLAT concepts. The textual interface, by itself, simplifies running L-FLAT with Mooshak.

\section{Conclusions and Future Work}

We are currently inspecting, revising, and documenting L-FLAT. Our goal is to ensure a stable foundation for future development and to provide teachers and students with a reliable and friendly tool for teaching and learning Formal Languages and Automata Theory. We also hope to enlist other developers to help expand the L-FLAT feature set. 

Future work will include adding support for $\omega$-automata (using the support for coinduction on recent Logtalk versions) and other language mechanisms (e.g. attribute grammars), adding conversion to and from the XML format used by JFLAP in order to share languages and mechanisms definitions, adding more conversion operations between mechanisms (e.g. between context-free grammars and pushdown automata), expanding the current set of examples, and exploring possible options for developing a graphical user interface (a task hindered by the lack of Prolog standards for interfacing with graphical  user interface toolkits).

We are also considering taking advantage of the support for \textsl{tabling} present on some Logtalk-compatible Prolog compilers. We are confident that we can use Logtalk support for conditional compilation in order to provide an implementation fallback for Prolog compilers that do not support tabling or where tabling support is an optional feature that might not be present at runtime.

\paragraph*{Authors' Contributions.} PM is the developer of the Logtalk programming language and a co-developer of L-FLAT. AMD is a co-developer of the original P-FLAT toolkit and a co-developer of L-FLAT.

\paragraph*{Acknowledgements.} We thank Miguel Wermelinger, who had the original idea and was co-author of the 2005 version of P-FLAT, and Luís Monteiro, whose teaching materials largely influenced the current features of L-FLAT. We also thank Paul Crocker for feedback and suggestions regarding L-FLAT features.

\bibliographystyle{splncs}
\bibliography{lflat,pmoura}

\newpage
\appendix

%
%
%

\section{L-FLAT Architecture}
\label{lflatarch}

\begin{figure}[ht!]
	\begin{center}
		\includegraphics[scale=0.65] {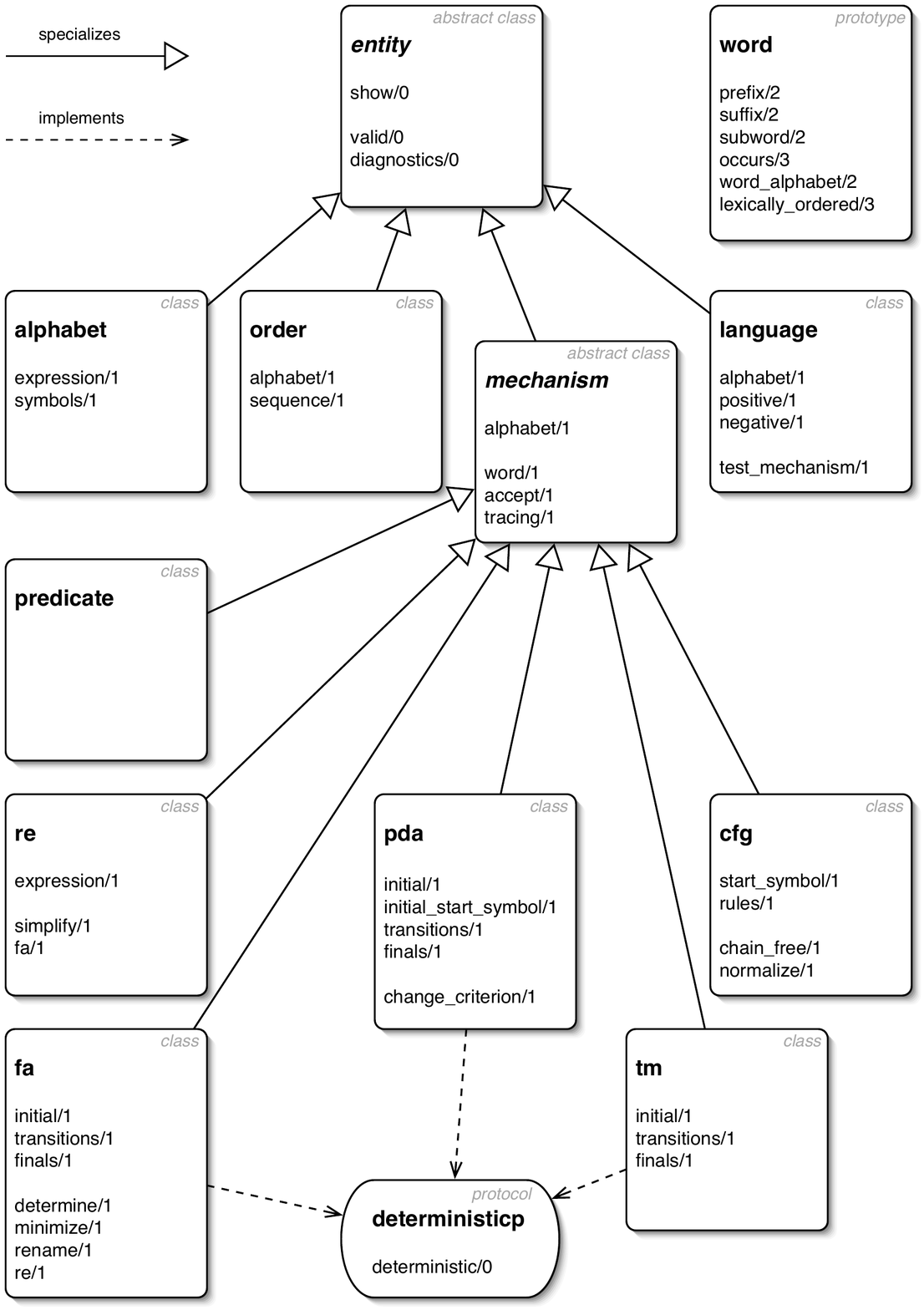}
		\caption{Main L-FLAT entities and main predicates}
		\label{arch}
	\end{center}
\end{figure}

\end{document}